# The 2$^{nd}$ Generation VLTI path to performance

Julien Woillez [*a],
Jaime, Alonso [a], Jean-Philippe Berger [a], Henri Bonnet [a], Willem-Jan de Wit [a],
Sebastian Egner [a], Frank, Eisenhauer [b], Frédéric Gonté [a], Sylvain Guieu [a],
Pierre Haguenauer [a], Antoine Mérand [a], Lorenzo Pettazzi [a],
Sébastien Poupar [a], Markus Schöller [a], Nicolas Schuhler [a].

[a] European Southern Observatory, Karl Schwarzschildstr. 2, Garching 85748, Germany.
[b] Max-Planck-Institut für extraterrestrische Physik, Giessenbachstraße, Garching 85746, Germany.

## ABSTRACT

The upgrade of the VLTI infrastructure for the 2$^{nd}$ generation instruments is now complete with the transformation of the laboratory, and installation of star separators on both the 1.8-m Auxiliary Telescopes (ATs) and the 8-m Unit Telescopes (UTs). The Gravity fringe tracker has had a full semester of commissioning on the ATs, and a first look at the UTs. The CIAO infrared wavefront sensor is about to demonstrate its performance relative to the visible wavefront sensor MACAO. First astrometric measurements on the ATs and astrometric qualification of the UTs are on-going. Now is a good time to revisit the performance roadmap for VLTI that was initiated in 2014, which aimed at coherently driving the developments of the interferometer, and especially its performance, in support to the new generation of instruments: Gravity and MATISSE.

**Keywords:** Very Large Telescope Interferometer, Optical Interferometry, Adaptive Optics, Fringe Tracking, Performance.

## 1. INTRODUCTION

In early March 2015, the Very Large Telescope Interferometer (VLTI) was shut down for a period of ~7 months. This was necessary to perform a functional upgrade of the infrastructure and prepare the arrival of the next generation instruments. These changes are presented in detail by Gonté et al. [1], and involve: the construction of a new maintenance station for the Auxiliary Telescope (AT), to facilitate the future installation of NAOMI [2] on the ATs and support the increased uptime requirement from the 4T instruments (PIONIER [3], Gravity [4], MATISSE [5]), the decommissioning of MIDI [6] and PRIMA FSU [7] and relocation of PIONIER [3] in order to make room for Gravity [4] and MATISSE [5], the upgrade of the Interferometric and Combined Coudé laboratories with new cooling, power, network, and cryogenic supply lines, the installation of star separators on AT1 and AT2 (AT3 and AT4 had already been equipped by the PRIMA facility project [8]). This transformation of the AT array was completed in time for the installation of the Gravity beam combiner in October 2015. Along with an equivalent cooling, power, network, and cryogenic supply line upgrade of the outer Coudé rooms, the installation of star separators on the Unit Telescopes (UT) was completed a few months later, in January 2016, in time for the arrival of the first infrared wavefront sensor CIAO [9] on UT1, in March 2016. UT4 was then equipped in May 2016; UT2 and UT3 will follow by September 2016. Finally, the over 10-years-old visible adaptive optics system, MACAO [10], was upgraded to address obsolescence issues and improve its performance. This activity is presented in more details by Haguenauer et al. [11]. This paper focuses on the performance of VLTI, first with the Auxiliary Telescopes (section 2), then with the the Unit Telescopes (section 3).

---

[*] E-mail: jwoillez@eso.org

## 2. AUXILIARY ARRAY PERFORMANCE

**2.1 Costs and benefits of the Star Separators on the ATs**

Despite the cancellation of PRIMA astrometry, the installation of star separators on the ATs remained a requirement for the Gravity instrument: it needs the 4" field of view that only the Star Separators can deliver. The previous single field interface, without proper relay of the telescope pupil to the entrance of the delay lines, and delivering a beam with a diameter of 18 mm (compression ratio of 100 = 1.8 m / 18 mm), was not capable of this. The limited field of view of the single field was also responsible for severe transmission loss impacting observations with stations far from the VLTI laboratory. With the Star Separators come variable curvature mirrors similar to the one already found at the focus of the delay lines [12]. Combined with a lower beam compression ratio of 22.5 (1.8 m / 80 mm), a 4" field of view is transmitted from all the stations of the AT array.

However, this transmitted field of view improvement comes at a transmission cost, due to the additional 11 reflections introduced by this system (to be exact: K-mirror, star separator, and beam compressor contains 3+8+3 = 14 reflections minus the 3 reflections of the old single-field system). Following the upgrade, the transmission of the AT array was verified in the near infrared as shown in Figure 1. The relative transmission was measured at ~80% (what is expected from 11 reflections at 98% each), except for AT4 where the loss is more pronounced with a transmission at ~30% (11 reflections at 89% each). The current hypothesis for the poor AT4 transmission is a degraded telescope transmission, rather than the many reflections in the star separator itself.

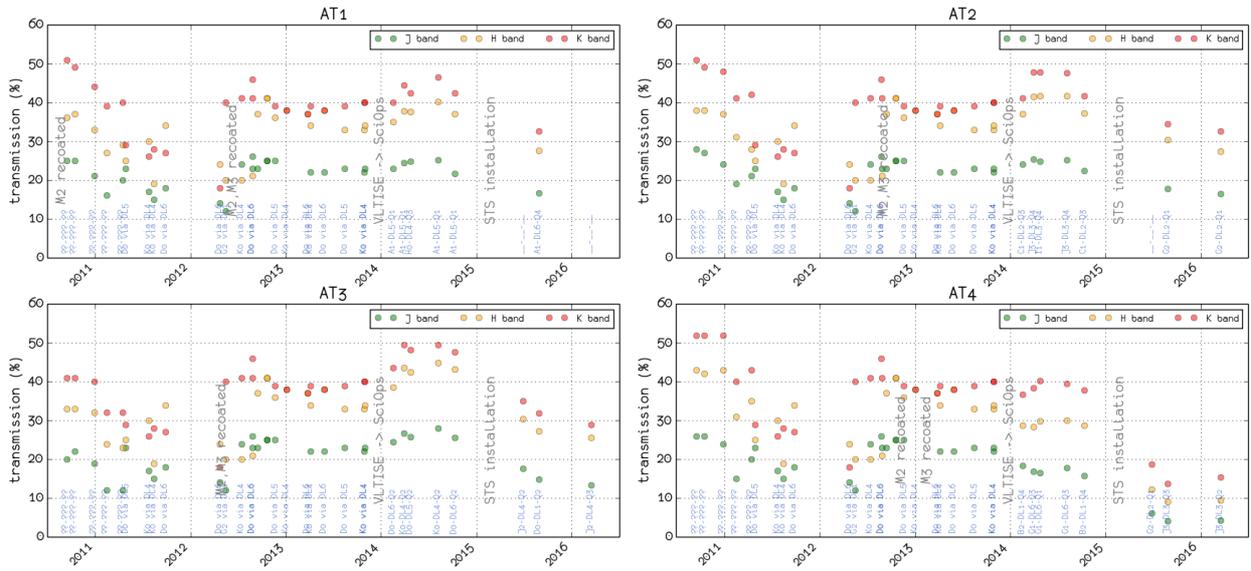

**Figure 1: Evolution of the infrared transmission of the ATs since 2011. The star separators where installed in the measurement gap at the beginning of 2015.**

The transmission loss resulting from the star separator upgrade triggered an investigation of the transmission of the optics upstream. The Coudé train and telescope optics were found significantly degraded. The M4 to M8 mirrors are the exact same telescope optics with which first lights was achieved over 10 years ago [13]. Table 1 summarizes the impact on the visible transmission (AT4 being the worst), which is at an unacceptable level for NAOMI [2], the future AO system of the ATs (see section 2.3). In consequence, a complete replacement of the Coudé train has been scheduled between the end of 2016 and the beginning of 2017.

Table 1: Measured transmission of the AT and UT telescopes in visible (V) to the Coudé focus, and in the infrared (J, H, and K) to the VLTI laboratory. The transmission is generally lower on the ATs than on the UTs due to the additional 4 reflections (+3 from the K mirror and +1 from the STS) and the significant AT Coudé train degradation. An equivalent measurement of the UT transmission in the visible, affecting the visible AO system MACAO, is pending.

|   | AT1 | AT2 | AT2 | AT4 | UT1 | UT2 | UT3 | UT4 |
|---|---|---|---|---|---|---|---|---|
| V | 1.5% | 2.3% | 1.8% | 0.4% |   |   |   |   |
| J | 11% | 13% | 12% | 4% | 17% | 18% | 14% | 18% |
| H | 17% | 23% | 21% | 10% | 33% | 37% | 28% | 33% |
| K | 26% | 29% | 25% | 15% | 36% | 39% | 32% | 33% |

The first element of the Star Separators of the ATs is a K-mirror responsible for compensating the field rotation at the Coudé focus and adjusting it onto the M10 mirror which splits this field into two halves A and B. Since PRIMA astrometry, the polarization retardance of this component is known to have a detrimental impact on interferometric observations, when combining telescopes located North and South of the delay lines tunnel [14]. Between Northern and Southern ATs, the field rotation at the Coudé focus is different by 180 deg, which is compensated by the K-mirror with a physical rotation offset of 90 deg, which, in turn, generates the detrimental differential polarization retardance. Luckily, single field instruments do not need this field rotation to be compensated (and indeed, it was not necessary when the single field relay optics were in use). Special modifications were therefore put in place, at the level of the Instrument Supervisory Software (ISS), to achieve the same K-mirror orientation, when single targets are observed.

Finally, the new star separators have a few other advantages over the single field. They have fast field and pupil actuators. By delivering 4.444 times larger beams (80 mm over 18 mm), the pupil instabilities associated to the delay line motions are reduced by the same amount.

This situation led to the decision to retire the single field interface between the ATs and VLTI, keeping the star separators as the only interface. This decision simplifies the operations of VLTI (there is no need to reconfigure), and the development of NAOMI [2] (to only support the star separators).

## 2.2 Vibration hunt with the Gravity Fringe Tracker

To assess the performance of VLTI, there is no better tool than the fringe tracker of Gravity. It represents a significant improvement over the two previous fringe trackers FINITO and PRIMA FSU [7]: the frame rate is ~1 kHz, it combines all pairs among 4 telescopes, and provides feedback on the instantaneous Strehl and Optical Path Difference. The analysis below relies on the output of the Gravity pipeline [15] applied to the standard Gravity data products.

Shortly after the re-opening of VLTI in October 2015, a number of issues with vibrations on the ATs became apparent in the residuals of the newly installed Gravity fringe tracker. The positive y-axis of each plot in Figure 2 shows the situation in January 2016, after having diagnose the situation, but before any fix. AT3, the worst of all four ATs, display significant flux dropouts, and a residual optical path fluctuation of ~265 $nm_{RMS}$, dominated by vibrations.

AT3 was severely affected in injected flux and piston by a vibration at 57.5 x n Hz. This one was traced down to the presence of a faulty cooling fan in the electronic of the acquisition camera installed just below the star separators. The replacement of the fan and the installation of vibration dampers between this electronics box and the structure supporting the star separators solved this issue.

In addition, an issue dubbed "Ultra-Fast Flux Dropouts", showing in Figure 2 at n x 50.0 Hz, was still affecting the ATs. It had already been observed before the star separator upgrade, by PIONIER, FINITO, and PRIMA FSU [7]. Due to the high frequency content of this perturbation, the origin had to be related to a fast actuator, which was found to be the M6 piezo-actuated mirror, responsible for the fast tip/tilt correction. Further investigations narrowed down the issue to a grounding problem of the piezos, which was finally resolved on all four ATs by March 2016.

The negative sides of the plots in Figure 2 show the performance achieved after the two fixes above. Very little vibrations are left. The atmospheric perturbation is the dominating contribution to the fringe tracking residuals. In the median seeing conditions that corresponds to Figure 2, the fringe tracker loop rate could certainly be reduced at the expense of a limited increase of the optical path difference residuals. For nights with good seeing and low wind, the achieved optical path difference, with the Gravity fringe tracker operating at ~1 kHz, is on the order of 100~150 nm$_{RMS}$.

Unfortunately, the telescope performance degrades significantly at high wind speeds. As shown in Figure 3, when the wind speed is above ~8 m/s, two resonant frequencies of the M2 and M3 structures are excited and generate piston perturbations at ~25 Hz and ~65 Hz. To remedy these telescope structure resonances will be challenging.

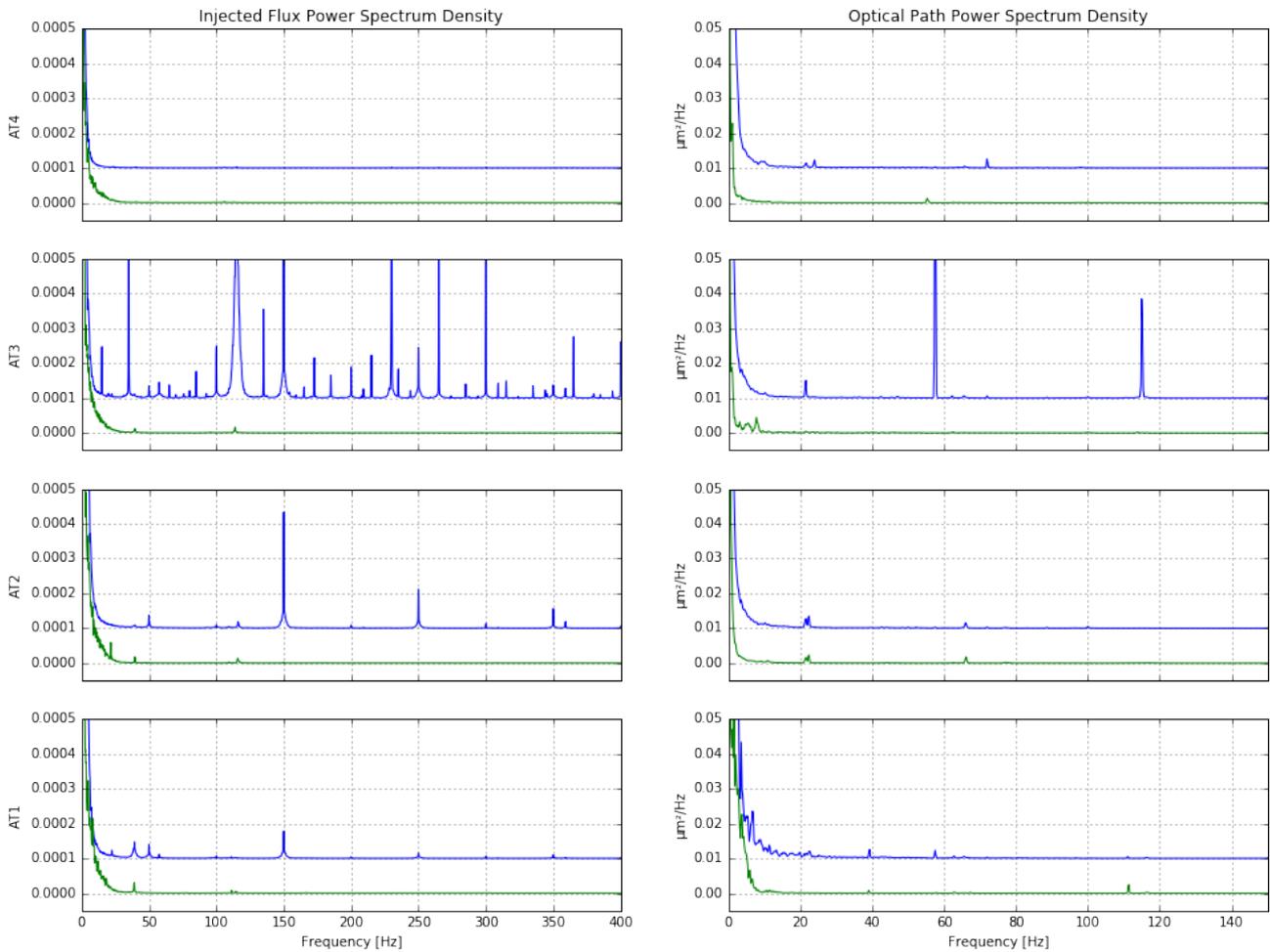

**Figure 2: Evolution of the Injected Flux (left) and Optical Path (right) power spectrum densities measured by the Gravity fringe tracker before (blue) and after (green) the vibration mitigation campaign on the Auxiliary Telescopes. The densities before vibration mitigation have been shifter up for clarity.**

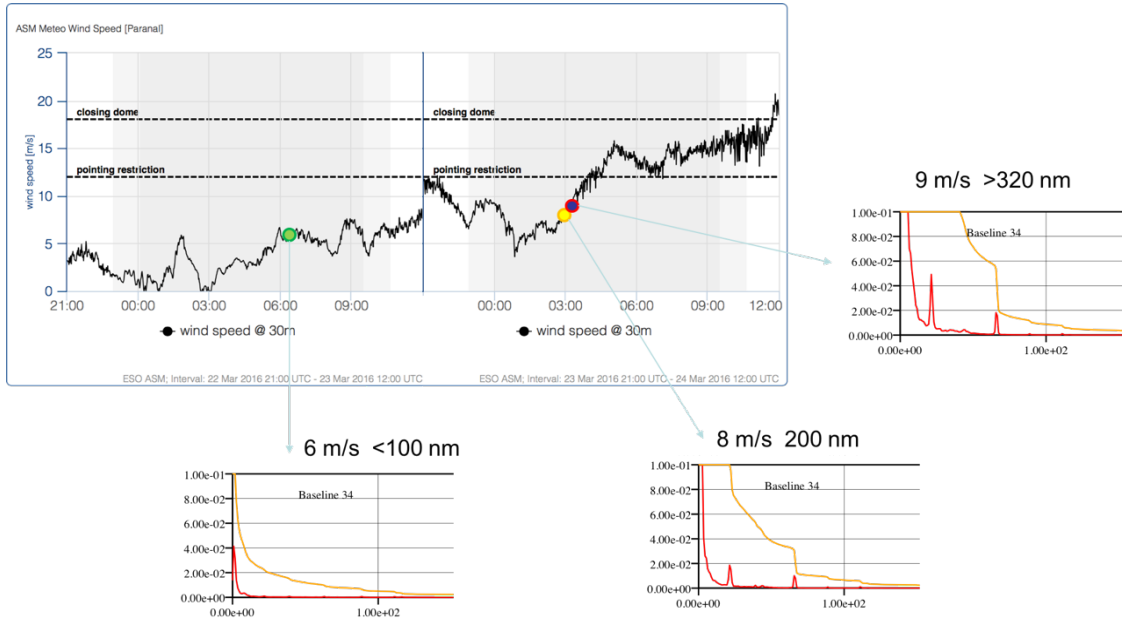

Figure 3: Auxiliary telescopes are sensitive to wind-shake. When the wind speed goes beyond ~8 m/s, disturbances at 25 Hz and 65 Hz start degrading the Gravity fringe tracker performance. Top left: wind speed versus time. Others: power spectrum densities (red) and reverse cumulated (orange) of the optical path difference residuals.

### 2.3 The next step: NAOMI

Despite the progress and achievements highlighted in the previous section, the auxiliary telescopes still deliver a poor Strehl. At the moment, this is the dominating factor in terms of performance. The primary mirror of the tip/tilt corrected auxiliary telescopes has been chosen to optimize the sensitivity of the array rather than support fringe tracking. The best theoretical illustration of this can be found in Figure 6 of Tatulli et al. [16], where the statistics of the instantaneous Strehl of an 1.8 m AT with tip/tilt correction shows a significantly high probability for very low Strehl events. In comparison, the AO correction of MACAO on an 8 m UT results in a much narrower distribution. Stabilizing the Strehl away from zero is the primary objective of the NAOMI project [2], in addition to improving performance when the seeing degrades.

When the wind speed reaches below 3 m/s, most of the ground layer turbulence is suppressed. This ground layer is responsible for most of the effective turbulence degradation observed on the ATs in comparison to the UTs [17]. However, these low wind nights, rather than being associated with excellent seeing, suffer from a form of local turbulence named "boiling". The telescope point spread function, as seen on e.g. the IRIS tip/tilt sensor [18] in the VLTI laboratory, contains multiple slowly-evolving speckles. The fringe tracking performance is severely impacted. The NAOMI [2] adaptive optics will be able to correct this turbulence very efficiently. The best Paranal nights, with low wind and excellent seeing, will then be accessible by VLTI. These are the nights when the Gravity fringe tracker could be significantly slowed down and thus VLTI will be able to reach significantly fainter objects.

The NAOMI adaptive optics, presented in greater details by Gonté et al. [2], are expected to be installed on all four ATs by late 2018.

# 3. UNIT ARRAY PERFORMANCE

Known to suffer more from vibrations, the UTs have been the focus of intense vibration mitigation efforts over the years [19],[20]. These efforts have continued over the past two years, motivated by the arrival of the Gravity instrument, and its challenging Galactic Center science case.

## 3.1 The impact of the telescope secondary mirror

As reported two years ago by both VLTI [20] and SPHERE [21], the field stabilization control loop, which runs between the guide probe and the UT secondary mirror, was responsible for significant degradation of the wavefront delivered to high angular resolution instruments. The guide probe would send tip/tilt guiding offsets that would be instantly corrected by the high control bandwidth secondary mirror. From the AO systems, these corrections would appear as step functions, with a frequency content beyond their correction capability. Replacing this field stabilization by a gentler auto-guiding loop to the telescope axes fixed the issue.

However, this was not the end of the M2 troubles. Non-atmospheric broad-band perturbations up to 100 Hz were still present in the adaptive optics telemetries. These were again traced to the same field stabilization actuator at the level of M2. This actuator has a local control loop that is aggressively tuned to fulfill the 5 Hz and high duty cycle chopping requirements, and introduces tip/tilt noise up to ~100 Hz at a level that is only significant for AO corrected instruments like SPHERE and VLTI. When the local control loop of the M2 tip/tilt stage is opened, the perturbation is removed.

Under the assumption that low Strehl events have the largest impact on the fringe tracking limiting magnitude, Figure 4 clearly shows the positive impacts of these two improvements on the MACAO and VLTI performance.

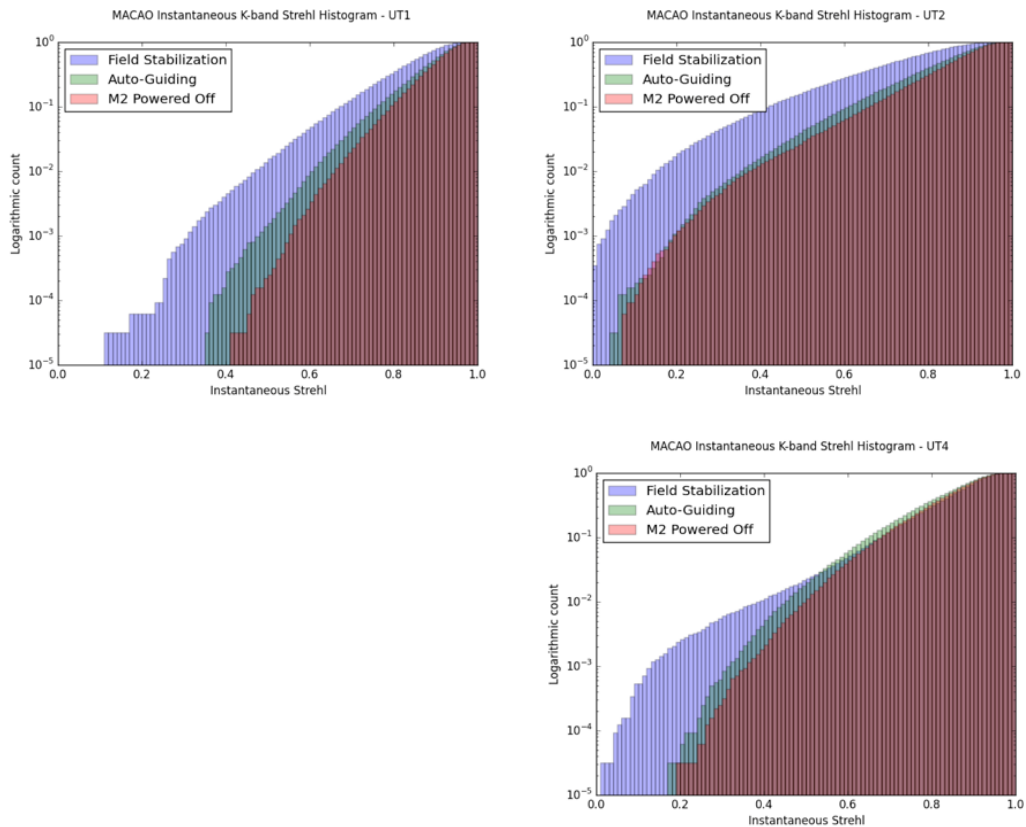

**Figure 4: Improvements of the instantaneous Strehl delivered by the MACAO adaptive optics, after the M2 related fixes.**

## 3.2 MACAO performance upgrades

The 10 years old adaptive optics systems MACAO was upgraded to address some obsolescence issues. We report two significant improvements: an increase of the loop rate from 420 Hz to 1050 Hz, and the implementation of a vibration tracking algorithm in the control loop. The functional and performance aspects of this upgrade are extensively detailed in Haguenauer et al. [11]. As such, we simply illustrate the improvements in Figure 5 below. We however remind the reader that both features improve performance only when observing bright objects.

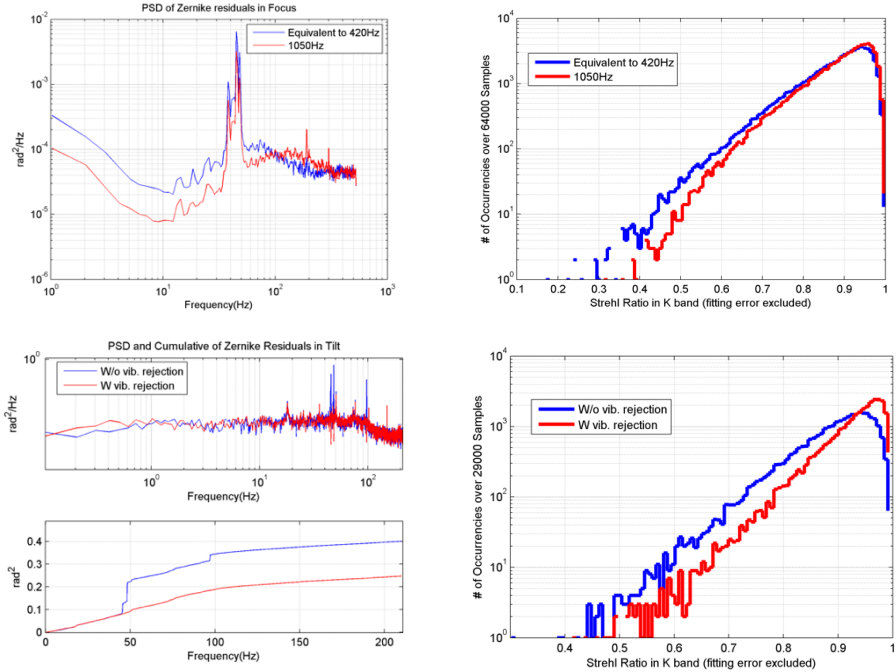

**Figure 5: Typical performance improvements associated to the MACAO upgrade. Top: increase of the loop rate from 420 Hz to 1050 Hz. Bottom: implementation of a vibration tracking algorithm inside the control loop. Left: power spectral densities. Right: instantaneous Strehl histograms.**

## 3.3 The Unit Telescopes seen by the Gravity Fringe Tracker

The Gravity Fringe Tracker achieved proper first light with the four UTs in March 2016, with a repeat in May. As with the ATs, the fringe tracker telemetry gives a clear picture of the remaining performance issues to be addressed. As shown in Figure 6, with respect to optical path vibrations, the main offenders are the 96 Hz of the auxiliary cooling pumps on all four UTs, the 72.5 Hz of the LGSF cooling pump on UT4, and a 47.5 Hz of a still unknown origin on UT3 (most prominently present at the level of M2 and observed by at least two other UT3 instruments: VISIR and VIMOS). Similar vibration frequencies are visible in the injected flux, with more prominent features in the 45 Hz ~ 50 Hz range probably associated to the diverse cooling fans present in all UT electronic cabinets.

## 3.4 Vibrations mitigation and control

Plans are already in place to address the 72.5 Hz on UT4 by replacing the aging LGSF cooling pumps by a quieter model. This replacement has already been tested and proven to reduce this vibration as observed by the telescope accelerometers. With respect to the 96 Hz, replacement tests are scheduled to determine, before the end of the year, if a similar auxiliary pump replacement would reduce this vibration. Not much is known about 47.5 Hz on UT3, beside the facts that it appears intermittently and more strongly at the level of its M2; further investigations are on-going. From the perspective of the optical path, if all four vibrations were addressed (47.5 Hz on UT3, 72.5 Hz on UT4, and 96 Hz on all UTs), the residuals at 5 Hz would fall below 180 nm, assuming that the residuals are mostly continuum (see Note 1 in the legend of Figure 6).

While this investigation/mitigation campaign is on-going, we have started exploring, as a fallback solution, a laser-based vibration metrology covering the optical path from the laboratory to the secondary mirror of the UTs. The current concept is largely inspired from the PRIMET concept [22], with the addition of 1) an optical component to make it polarization insensitive, and 2) an additional fast actuator at the level of the beam compressor to close the loop with the highest rejection factor possible.

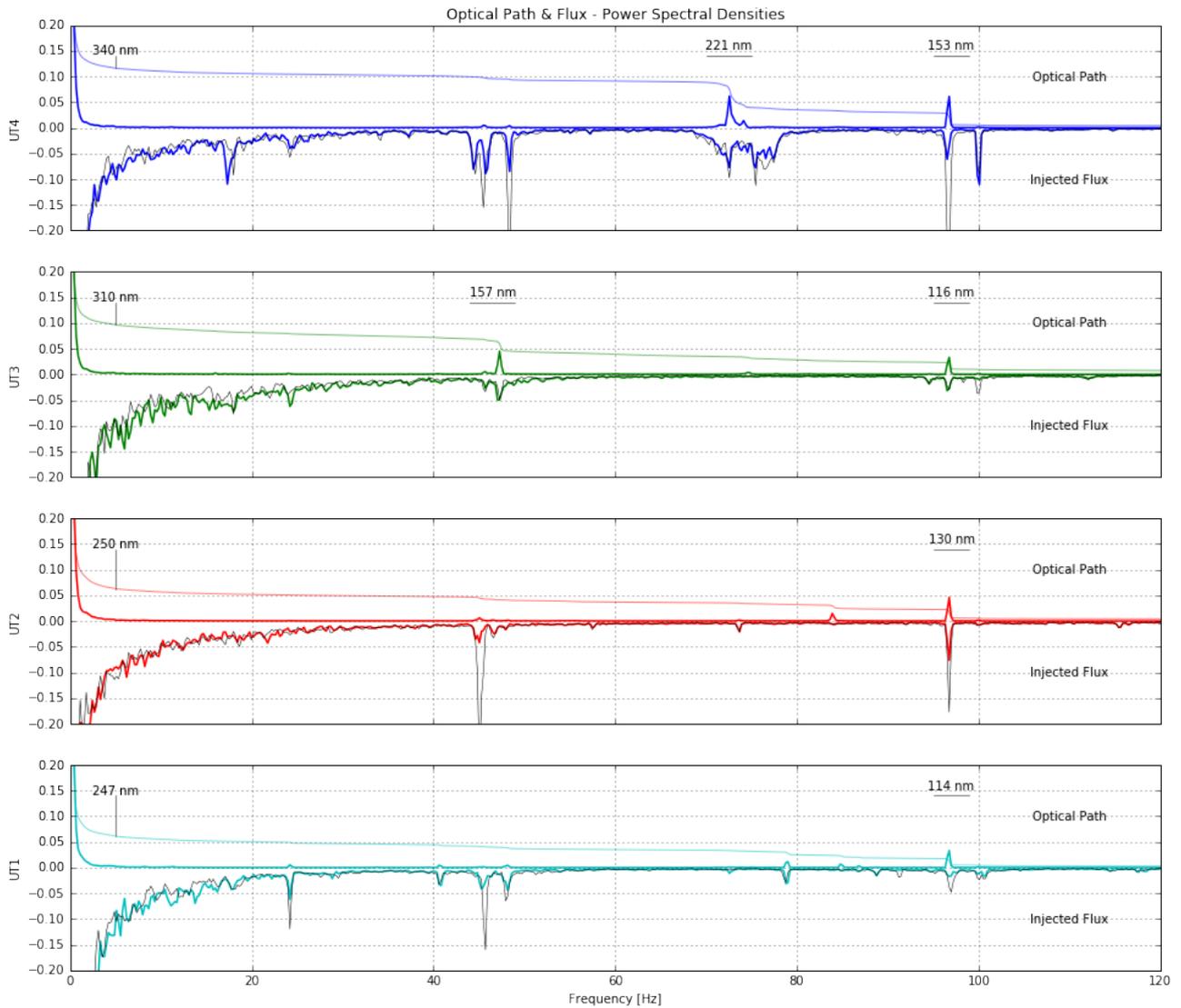

**Figure 6: Unit Telescopes vibrations observed by the Gravity Fringe Tracker. Positive: Optical Path Power Spectral Density (thick line) and reverse cumulated Power Spectrum (thin line), annotated with vibration contributions and cumulated disturbance at 5 Hz. Negative: Injected Flux Power Spectral Density before (thin line) and after (thick line) AO vibration tracking correction.** Note 1: For each telescope, the optical path power spectral density is estimated as the minimum at each wavelength of the optical path difference power spectral densities of the three baselines the given telescope participates to. This approach is correct in identifying vibrations unique to a given telescope, but overestimates the continuum by a factor $\sqrt{2}$. Note 2: For the Optical Path, the control loops between the accelerometers on M1-M2-M3 and the VLTI delay lines are active and cancels the vibrations up to ~35 Hz. Contrary to the vibration tracking loop inside MACAO, which only works for bright objects, the accelerometer loop is functional all the time. As such, it is fair to simply hide these corrected vibrations from the vibration footprints of the UTs.

## 4. CONCLUSION

Since 2014, the upgrade of the infrastructure became the focus of the VLTI project. For the past two years, significant performance improvements have been achieved on both AT and UT arrays, bringing VLTI many steps closer to a situation compatible with robust fringe tracking. Despite the progress, many challenging upgrades are still ahead of us and need to be completed well within the next 2 years, before the rendezvous of S2 with our galactic center black hole in 2018.